# Upper critical magnetic field in $K_{0.83}Fe_{1.83}Se_2$ and $Eu_{0.5}K_{0.5}Fe_2As_2$ single crystals

Vitaly A. Gasparov[1], A. Audouard[2], L. Drigo[2], A.I. Rodigin[3], C.T. Lin[4], W.P. Liu[4], M. Zhang[5], A.F. Wang[5], X.H. Chen[5], H.S. Jeevan[6], J. Maiwald[6], and P. Gegenwart[6]

[1] *Institute of Solid State Physics RAS, 142432, Chernogolovka, Moscow District, Russian Federation*
[2] *Laboratoire National des Champs Magnétiques Intenses (UPR 3228 CNRS, INSA, UJF, UPS) 143 avenue de Rangueil, F-31400 Toulouse, France*
[3] *Lomonosov Moscow State University, GSP-1, Leninskie Gory, Moscow, 119991, Russian Federation*
[4] *Max Planck Institute for Solid State Research, 70569 Stuttgart, Germany*
[5] *Hefei National Laboratory for Physical Science at Micro scale and Department of Physics, University of Science and Technology of China, Hefei, Anhui, 230026, People's Republic of China*
[6] *I. Physik. Institut, Georg-August-Universität Göttingen, D-37077 Göttingen, Germany*

**Abstract**

The *H-T* phase diagrams of single crystalline electron-doped $K_{0.83}Fe_{1.83}Se_2$ (KFS1), $K_{0.8}Fe_2Se_2$ (KFS2) and hole-doped $Eu_{0.5}K_{0.5}Fe_2As_2$ (EKFA) have been deduced from tunnel diode oscillator-based contactless measurements in pulsed magnetic fields up to 57 T for the inter-plane (*H//c*) and in-plane (*H//ab*) directions. The temperature dependence of the upper critical magnetic field $H_{c2}(T)$ relevant to EFKA is accounted for by the Pauli model including an anisotropic Pauli paramagnetic contribution ($\mu_B H_p = 114$ T for *H//ab* and 86 T for *H//c*). This is also the case of KFS1 and KFS2 for *H//ab* whereas a significant upward curvature, accounted for by a two-gap model, is observed for *H//c*. Despite the presence of antiferromagnetic lattice order within the superconducting state of the studied compounds, no influence of magnetic ordering on the temperature dependence of $H_{c2}(T)$ is observed.

## I. Introduction

Since the discovery of superconductivity in the *FeAs* and *FeSe* - based families, intensive studies have focused on the dimensionality of these superconductors (see references in [1,2]). Although the Fermi surfaces (FS) are quasi-two-dimensional [1,2], reports on the anisotropy of the upper critical field, $H_{c2}(T)$, are quite puzzling (see references in [1-10]). In the field range below *10 T*, where $H_{c2}(T)$ is limited by orbital pair breaking, $H_{c2}(0)$ can be evaluated through the slope of $dH_{c2}/dT|_{Tc}$ close to $T_c$ according to the well-known Werthamer–Helfand–Hohenberg (WHH) model for the orbital critical magnetic field $H^{orb}_{c2}(0) \approx -0.69T_c (dH_{c2}/dT)|_{Tc}$ [11]. While a significant anisotropy of $\gamma = H^{ab}_{c2}(0)/H^c_{c2}(0)$ is reported for (1111) and (122) iron pnictides in this temperature range, direct measurements of $H_{c2}(T)$ in pulsed magnetic fields have shown that the actual anisotropy of $H_{c2}(0)$ for an electron-, and a hole-doped, 122 superconductor becomes very small at low temperatures [3-10]. While in $Ba_{0.68}K_{0.32}Fe_2As_2$ this anisotropy is washed out by Pauli spin paramagnetism [5], a two-band model must be invoked to account for the behavior observed in $Ba(Fe_{0.93}Co_{0.07})_2As_2$ [6] and $Sr_{1-x}Eu_x(Fe_{0.89}Co_{0.11})_2As_2$ [9] for *H//c*.

More recently, a new class of Fe chalcogenide-based superconductors: $A_xFe_{2-y}Se_2$ (A =K, Rb, Cs, Tl) with $T_c$ above 30K were reported [12-17]. Many differences between Fe-pnictide and Fe-chalcogenide are observed: (i) At variance with Fe-pnictide superconductors, the FS of which involves hole pockets [1,2], ARPES data reveal the existence of only two electron like bands at the $M(\pi,0)$ point and around the Brillouin zone centre $\Gamma(0,0)$ in $A_xFe_{2-y}Se_2$ compounds, both of them having nearly isotropic superconducting gap [18-22]. Therefore, since a $S_{+/-}$ paring symmetry is expected when both hole-like and electron-like pockets are present [23], the absence of hole-like pocket at $\Gamma$ makes this hypothesis rather questionable; (ii) The hole-like bands near the zone center $\Gamma$ are shifted down below $E_F$ and thus do not contribute to the FS [18–22]. (iii) In contrast to the metallic-like behavior of the Fe-pnictide superconductors, the resistivity increases as the temperature decreases from room temperature with a broad hump at 100 – 200 K. It is followed by a metallic-like behavior at lower temperatures with a superconducting transition ($T_c$=29-33 K) observed for a wide range of concentrations (0.6< x <1 and 0<y <0.59); (iv) Magnetic susceptibility, resistivity and neutron diffraction data evidence an antiferromagnetic (AFM) transition with Néel temperature $T_N$ as high as 500 K to 540 K, depending on the composition, for $AFe_{2-y}Se_2$ (A = $K_{0.8}$, $Rb_{0.8}$, $Cs_{0.8}$, $Tl_{0.4}K_{0.3}$ and $Tl_{0.4}Rb_{0.4}$) [24,25,26]. In contrast to this feature, an electron spin resonance (ESR) signal arising from paramagnetic Fe ions is detected at room temperature for both $K_xFe_{2-y}Se_2$ and $K_xFe_{2-y}Se_{1.6}S_{0.4}$ crystals [27]. Upon cooling, the intensity of the ESR spectrum abruptly disappeared below 140 K. It can therefore be concluded that a transition from paramagnetic to AFM state takes place at T<140, as reported in Ref. [27]. As a consequence, the coexistence of superconductivity and lattice AFM at $T<T_c$ must be considered.

It is of interest to study the temperature dependence of the anisotropic upper critical field, $H_{c2}(T)$, in Fe chalcogenides in order to determine whether or not their behavior is similar to or different from those observed in the FeAs-based 122-type compounds [5,6]. More specifically, the question is to determine if the Fe- selenide superconductors give rise to a new type of superconductivity due to coexistence of AFM and superconductivity, or remain similar to paramagnetic Fe - pnictides. So far, the bulk $H_{c2}(T)$ for $K_{0.73}Fe_{1.68}Se_2$ and $Rb_{1-x}Fe_{2-y}Se_2$ single crystals has been determined over a wide range of temperatures and magnetic fields by means of measuring either the electrical resistance, or the radio-frequency penetration depth in a pulsed magnetic field up to 60 T [7,8,9,28]. While the behavior of $H_{c2}(T)$ is very similar to that of several FeAs-based 122-type materials, a surface superconductivity was observed for $K_{0.73}Fe_{1.68}Se_2$ single crystals for $H // ab$ from magnetic susceptibility experiments [29]. Besides, no indication of third $H_{c3}$ was observed for KFS single crystals for $H // ab$ from $H_{c2}(T)$ in pulsed field experiments even though the superconducting transition at $H // ab$ is strongly broadened [8]. Furthermore, coexistence of lattice AFM due to short range magnetic ordering of the $Eu^{2+}$ ions and superconductivity was observed recently below 10 K for $Eu_{0.5}K_{0.5}Fe_2As_2$ (EKFA) polycrystalline samples [30, 31]. Since both KFS and EKFA exhibit coexistence of AFM and superconductivity, it is of interest to determine whether or not the temperature dependence of $H_{c2}(T)$ is influenced by AFM ordering [9].

Here, we report on the study of the temperature dependent upper critical magnetic field in the directions parallel and perpendicular to the crystallographic $c$ axis in electron-doped $K_{0.83}Fe_{1.83}Se_2$ (KFS1), $K_{0.8}Fe_2Se_2$ (KFS2) and hole-doped EKFA single crystals by radio-frequency tunnel-diode-oscillator technique. It is evidenced that for EKFA the temperature dependence of $H_{c2}(T)$ can be explained taking into account Pauli spin paramagnetism. The latter substantially limiting $H_{c2}(T)$ and, in turns, the anisotropy as the temperature decreases below $T_c$. In contrast, Pauli paramagnetic pair breaking is only relevant for $H // ab$ for KFS which exhibits a two gap behavior for $H // c$.

## II. Experimental

Superconducting KFS1 single crystals have been grown by the optical floating-zone (OFZ) technique as described elsewhere in detail [32]. Single crystals with flat black color shiny surfaces were obtained. KFS2 single crystals, which display mirror like surfaces with golden like color, were grown in Hefei by the conventional high temperature flux method [24]. The actual composition of these crystals as determined by various methods is $K_{0.8}Fe_2Se_2$. Both KFS1 and KFS2 were quickly losing superconducting transition after staying short time in air after cleavage which requires thorough handling as reported in the following. EKFA single crystals were synthesized using the self flux method, in which the crystals grow out of a FeAs flux [33]. This method yields large plate-like single crystals with a typical dimension of 40 mm$^2$. *X*-ray diffraction data revealed that the surface of all the studied crystals is normal to the c axis.

The KFS1 samples were plates with dimensions of about 1 × 1 × 0.2 mm$^3$. Their resistance was measured using a four-probe van der Pauw technique from room temperature down to *4.2 K* on samples cleaved in air. The contacts to KFS1 sample corners were prepared with conducting silver paste and *Au* wires. The contact resistance was ≈ *1 Ω*. The ac current was applied along the sample. At variance with KFS1, KFS2 samples were cleaved in a Glow Box in Argon atmosphere and placed inside hermetic sapphire ampoule with a diameter of 3 mm in between sapphire plates. This ampoule was closed by a Teflon cork in a Glow Box before measurements, whereas no cleavage was necessary for EFKA crystals.

The device for the radio frequency (RF) magnetic penetration depth measurements is a *LC*-tank circuit powered by a tunnel diode oscillator (TDO) biased in the negative resistance region of the current-voltage characteristic, as reported in Ref. [34]. The coils are made from copper wire (50 μm in diameter) wound around either a Kapton tube or the above mentioned sapphire ampoule and connected with a similar compensated coil. Our TDO device is working as super heterodyne at fundamental resonant frequency in the range 16-20 MHz at $T_c$. After signal amplification, mixing with a frequency about 1 MHz below the fundamental frequency and demodulation, the resulting output oscillator frequency shift, which can be approximated by $\Delta f = 1/(2\pi\sqrt{LC})$, lies in the MHz range as displayed in Figs.2 and 3.

The RF technique was used because it provides contactless and much more sensitive measurements than conventional four-point technique for low-resistance samples such as superconductors at low temperatures [6]. The samples were placed inside one of the coil constituting the counter-wound pair. This coil pair was aligned either parallel or perpendicular to the field direction. As a result, the filling factor remains the same for both field directions which provides easily resolvable frequency shift. As the magnetic field increases, the transition to the normal state is detected from the shift in the resonance frequency. The resulting frequency variation versus magnetic field is, at first order, proportional to the changes in magnetic penetration depth.

The experiments were performed at fixed temperatures in pulsed magnetic fields of up to *57 T*, with pulse-decay durations of *0.25 ms*, at the Laboratoire National des Champs Magnétiques Intenses of Toulouse (CNRS). The magnetic field was applied either along the *c* axis or in the *ab* plane. Even though the reported data are collected during the decaying part of the pulse, we have checked that they are in agreement with data taken at the rising part, although with a reduced signal-to-noise ratio in the latter case, which confirms that the data are not affected by sample heating during the pulse.

### III. Results and discussion

Fig. 1 shows the temperature dependence of the resistivity for freshly cleaved KFS1 single crystal. The resistivity increases as the temperature decreases below room temperature with a broad hump at *180 K* and a metallic-like behavior at lower temperatures. From the mid point in the resistive transition, $T_c = 32.5$ *K* is obtained. According to recent ARPES data, this crossover can be understood as a temperature-induced transition from a metallic state at low temperature to an orbital-selective Mott phase at high temperatures [35], in which few orbital's are Mott-localized while the other remain itinerant. As was shown from this ARPES study the KFS superconductors evolves into a state in which the $d_{xy}$ bands have diminished spectral weight as the temperature increases while the $d_{xz}/d_{yz}$ bands remain metallic [35]. How this model is consis-tent with ESR spectra transition [27] is not yet clear. As a matter of fact, no structural transition is observed at $T_{hump}$.

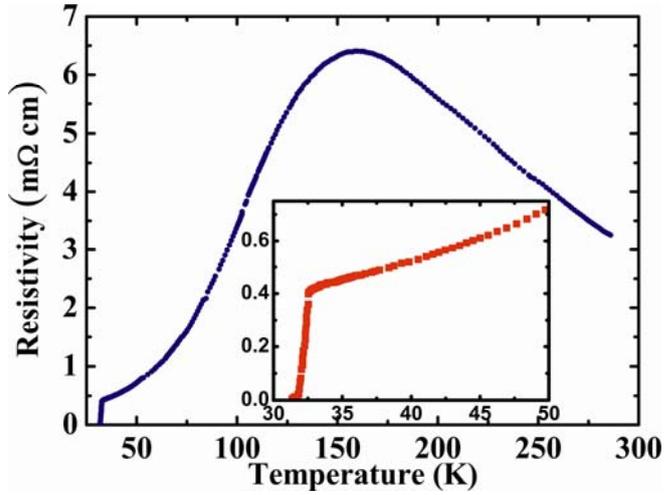

Fig.1 (Color online) Temperature dependence of the resistivity of KFS1 crystal

Fig. 2 and Fig. 3 display the field dependence at various temperatures of the TDO frequency for KFS2 and EKFA single crystals, respectively, in pulsed magnetic fields up to *57 T*, aligned parallel (Fig. 2a, Fig. 3a) and perpendicular (Fig. 2b, Fig. 3b) to the *ab* axis. Data for KFS1 sample are shown in the insets of Fig. 2a and Fig. 2b. The TDO data for KFS1 yield $T_c = 28$ K for *H// ab* (Fig. 2a) and *25 K* at *H // c* (Fig. 2b), respectively, which is lower than the value deduced from zero-field resistivity measurements (Fig. 1). We have checked that this shift is due to air degradation of the sample during its handling at room temperature. In contrast, we did not observed any shift of $T_c$ in the case of the KFS2 sample, since it was placed in a sapphire ampoule under argon atmosphere, (Fig. 2a and Fig. 2b). In line with the large $T_c$ values reported in Fig. 1, the studied compounds exhibit superconducting transitions up to very high fields, likely above 60 T at 0 K. The method for determining consistent $H_{c2}$ values from the data shown in Figs. 2 and 3 is based on identifying the point at which the steepest slope of the RF signal at the transition intercepts with the extrapolated normal-state background as discussed in Ref. [6].

As it is observed in the inset of Fig. 2a very large background is observed for KFS1 in the normal state, in particular for *H//ab* [8], which is otherwise almost flat for the KFS2 sample. The discrepancy in Δf between KFS1 and KFS2 is due to difference in filling factor, i.e. the ratio of the sample size to coil diameter which is larger in the latter case because the KFS2 crystal was placed in a sapphire ampoule. Besides, the concentration of the superconducting phase for KFS2 sample was larger due to both the absence dead surface layer and better crystal quality, as discussed below.

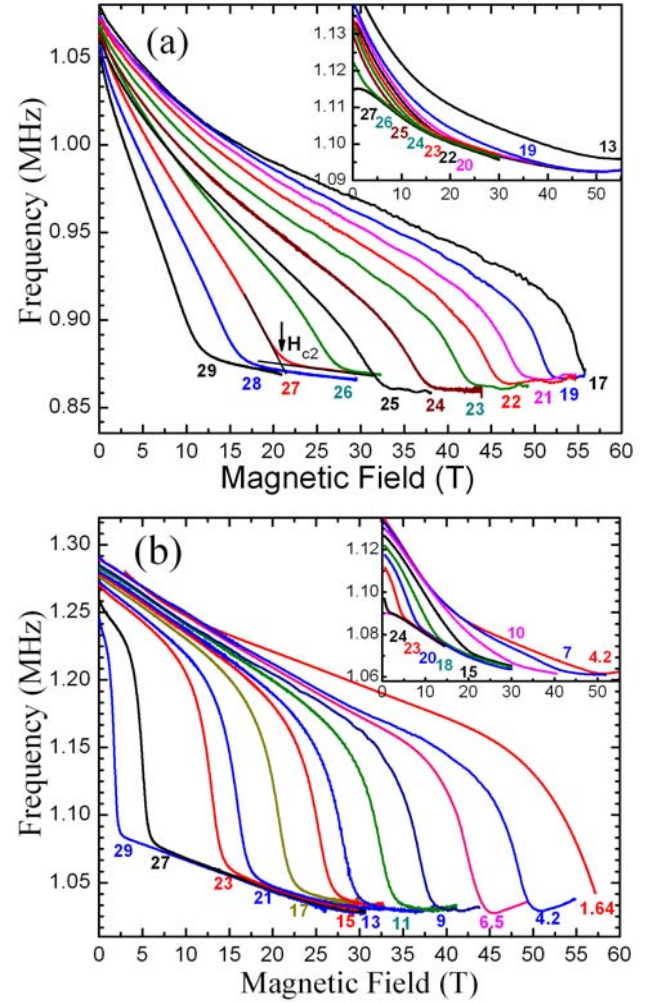

Fig. 2 (Color online) Field dependences of the TDO frequency shifts for KFS2 single crystal for magnetic fields applied: (a) along the *ab* direction at selected temperatures in Kelvin indicated on curves (inset shows similar data for KFS1 sample), (b) along the *c* direction (inset shows similar data for KFS1 sample). The arrow indicate $H_{c2}$ as the point deviating from background signal.

Approximating the background by a polynomial, the superconducting part of the signal can be extracted from the data. The superconducting transitions of EKFA single crystals in applied fields are pretty narrow, which is not the case for KFS1. The transition curves just move to higher fields with decreasing temperature for both field orientations. This feature made the determination of $H_{c2}(T)$ much easier in this latter case (see the construction lines in Fig. 2a and Fig.3b).

The resulting temperature dependence of $H^{ab}_{c2}$ and $H^{c}_{c2}$ for the hole-doped EKFA and electron-doped KFS1, KFS2 samples are shown in Fig. 4 and Fig. 5, respectively. As mentioned above, TDO data yield $T_c$ values for KFS1 lower than those deduced from resistivity measurements (see Fig. 1). For this reason, the normalized temperature (t = $T/T_c$) dependence is considered in Fig. 5. Close to $T_c$, the usual WHH linear temperature dependence of $H_{c2}$ is observed. The anisotropy parameter $\gamma = H^{ab}_{c2}(T)/H^{c}_{c2}(T)$, which is about 2 near $T_c$, decreases considerably at low temperature. Even though a very small anisotropy factor is also observed at low temperature for EKFA ($T_c$ = 31.5 K), a somewhat different behavior is observed in Fig. 4 since $H_{c2}(T)$ saturates both for H//ab and H//c.

As discussed in the following, the small anisotropy of $H_{c2}(T)$ observed for EKFA is due to a partial compensation of the orbital pair-breaking mechanism by Pauli paramagnetism, rather than to 2D FS effects. The temperature-dependent anisotropy is most likely due to these two independent pair-breaking mechanisms [5, 6, 36–39]: (i) at higher temperatures, Cooper pairing is suppressed by orbital currents that screen the external field, according to the well-known WHH model [11]; (ii) towards lower temperatures, the limiting effect is caused by the Zeeman splitting, i.e., when the Zeeman energy becomes larger than the condensation energy, the Pauli limit, $H_p$, is reached [36–39]. Indeed, assuming in a simple approximation, valid for weakly coupled BCS superconductors, that the superconducting gap is given by $2\Delta = 3.5k_BT_c$, $\mu_BH_p$ is $1.84 T_c$ [T/K] [39], resulting in $\mu_BH_p$ = 58 T for both KFS and EKFA. This paramagnetic limit is lower than the orbital limit, $H^*_{c2}(T)$, which is related to the slope $dH_{c2}(T)/dT$ close to $T_c$. Experimental data yield $dH^{c}_{c2}/dT$ = −1.68 T/K and $dH^{ab}_{c2}/dT$ = −5.5 T/K for KFS2, for H//c and H//ab, respectively, yielding, according to the WHH model [11], $\mu_BH^{*c}_{c2}(0)$ = 36 T and $\mu_BH^{*ab}_{c2}(0)$ = 120 T at T = 0. For EKFA, $dH^{c}_{c2}/dT$ = −3.5 T/K and $dH^{ab}_{c2}/dT$ = −5.3 T/K which result in higher estimates: $\mu_BH^{*c}_{c2}(0)$ = 75 T and $\mu_BH^{*ab}_{c2}(0)$ = 115 T at T = 0. The dotted lines in Fig. 4 and Fig. 5 display the temperature dependence of the orbital critical fields within the WHH approach for both field orientations and compounds ignoring the Pauli limit. These $H^*_{c2}(0)$ values, allow to derive the coherence lengths $\xi(0)$. We obtain $\xi_{ab}(0) = \sqrt{\phi_0/2\pi H^{*c}_{c2}(0)}$ = 2.83 nm and $\xi_c(0) = \phi_0/2\pi\xi_{ab}(0)H^{*ab}_{c2}(0)$ = 1.2 nm for KFS2, and $\xi_{ab}(0)$ = 2.1 nm and $\xi_c(0)$ = 1.36 nm for EKFA, respectively. Although anisotropic, the c-axis coherence lengths for EKFA, is nevertheless larger than the thickness of 0.32 nm of the conducting FeAs sheet indicating the 3D nature of the superconductivity for both compounds. Furthermore, when including spin Pauli paramagnetism, the upper critical field $H_{c2}(T)$ is reduced relatively to [36–38]:

$$H_{c2}(T) = \frac{H^*_{c2}}{\sqrt{(1+\alpha^2)}} \quad (1)$$

where $\alpha(T) = \sqrt{2}H^*_{c2}(0)/H_p(0)$ is the Maki parameter [37]. Fuchs proposed that $\alpha$ is temperature-dependent according to: $\alpha = \sqrt{2} H^*_{c2}(T)/H_p(0)$ [36]. The solid lines in Fig. 5 are the best fits for EKFA using Eq. (1) with temperature-dependent $\alpha$. However, exact equation for $H_p(T)$ with constant $\alpha$ as defined by Maki, is more complicated [38]:

$$\ln t + \text{Re}\{\psi[0.5 + 0.138\frac{h}{t}(1+i\alpha)] - \psi(0.5)\} = 0 \quad (2)$$

here t = $T/T_c$ , h = $H_p(T)/H^*(0)$, and $\psi(x)$ is the digamma function. The dashed lines in Fig. 4 are the best fits of Eq. (2) to the EKFA data, assuming constant $\alpha$. In these studies $H_{c2}(T)$ saturate both for H // ab and H// c. A very good agreement with the experimental data is observed for both field orientations within either the assumption of constant $\alpha$ or temperature-dependent for EKFA, albeit with a slightly different $H_p(0)$. Eq. (1) yields, as expected owing to the isotropic nature of the Pauli contribution, only one free parameter, namely $H_p$ = 114 T for both field orientations for EKFA while Eq. (2) yields anisotropic val-ues: $H^c_p = 114\ T$ ($\alpha = 0.85$) and $H^{ab}_p = 86\ T$ ($\alpha = 1.9$), for $H // ab$ and $H // c$, respectively. Almost the same value $H^{ab}_p = 114\ T$ is obtained from these fits for KFS2 for $H // ab$. Solid and dashed lines in Fig. 5 show the best fits for KFS2 of Eq. (1) and Eq. (2), respectively, for $H // ab$. These values are twice as large as the above estimate of $H_p = 58\ T$ for weakly coupled BCS superconductors. Nevertheless, this discrepancy is not unexpected since in the latter value neither many-body correlations nor strong-coupling effects are included [6,38].

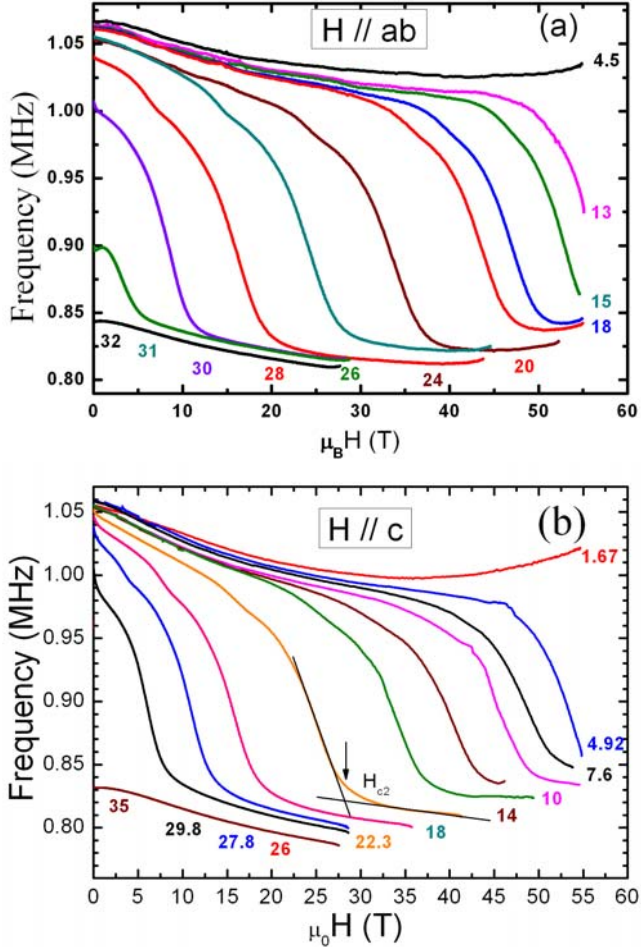

Fig.3. (Color online) Field dependences of the TDO frequency shifts for EKFA single crystal for magnetic fields applied: (a) along the ab direction at selected temperatures indicated on curves and (b) along the c direction. The arrow in (b) indicate $H_{c2}$ as the point deviating from background signal.

Unfortunately, we are not aware about direct data for the superconducting gap in EKFA from ARPES for comparison. Actually, an estimate of $\Delta$ from our data for $H_p$, using the Clogston equation $H_p = \Delta(0)/\sqrt{2}\mu_B$ [39] yields a superconducting gap value, $\Delta(0) = 9.3\ meV$ and thus a strong coupling value of 6.9 for $2\Delta/k_BT_c$. According to ARPES data of the $(K,Cs)_xFe_2Se_2$ compounds, the superconducting gap of the electron $\delta$ band around the $M$ point and $\kappa$ band at $\Gamma$ are about $10.3 - 8\ meV$ [18–22] and $4\ meV$ [18], respectively. The above estimated value is more in line with the gap of the $\delta$ band rather than that of the $\kappa$ band, the gap of which is twice as small.

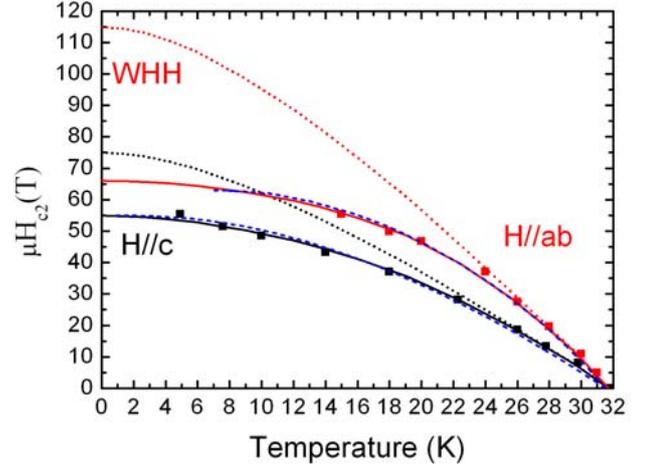

Fig. 4. (Color online) Temperature dependences of the upper critical fields, $H_{c2}(T)$, for EKFA single crystal for $H \| c$ (black squares) and $H \| ab$ (red squares). The dotted lines indicate the temperature dependences according to the WHH model neglecting Pauli-limiting for both samples. The solid lines show the dependences including Pauli pair breaking using Eq. (1) and dashed line from Eq.2.

A completely different behavior is observed for the nonstoichiometric electron-doped KFS1 and KFS2 ($T_c = 32.5\ K$) for $H // c$. At variance with the data for $H // ab$, the WHH model, cannot account for the data for $H // c$, even including the Pauli contribution (see Fig. 5). For this field orientation we observe a positive curvature at low temperature without any saturation, indicating that Pauli paramagnetic pair breaking is not essential for $H // c$. The upward curvature of $H^c_{c2}(T)$ can be accounted for by two-band features recently evidenced for $Ba(Fe_{0.93}Co_{0.07})_2As_2$ [6], and $Sr_{1-x}Eu_x(Fe_{0.89}Co_{0.11})_2As_2$ [9]. According to Gurevich [40], the zero-temperature value of $H_{c2}(0)$ is significantly enhanced in the two gap dirty limit superconductor model,

$$H_{c2}(0) = \frac{\phi_0 k_B T_c}{1.12\hbar\sqrt{D_1 D_2}} \exp\left(\frac{g}{2}\right) \quad (3)$$

as compared to the one-gap dirty-limit approximation $H_{c2}(0) = \phi_0 k_B T_c/1.12\hbar D$. Here, $g$ is a rather complicated function of the matrix of the BCS superconducting coupling constants $\lambda_{mm'} = \lambda^{ep}_{mm'} - \mu_{mm'}$, where $\lambda^{ep}_{mm'}$ are electron-phonon coupling constants and $\mu_{mm'}$ is the matrix of the Coulomb pseudopotential. In a simple approximation using the same inter-band, $\lambda_{12} = \lambda_{21} = 0.5$, and intra-band, $\lambda_{22} = \lambda_{11} = 0.5$, coupling constants [6], the equation for $H_{c2}(T)$ takes the simple Usadel form:

$$a_1[\ln t + U(h)] + a_2[\ln t + U(\eta h)] = 0. \quad (4)$$

Here, $a_1 = 1 + \lambda_-/\lambda_0 = 1$; $a_2 = 1 - \lambda_-/\lambda_0 = 1$; $\lambda_0 = (\lambda_-^2 + 4\lambda_{12}\lambda_{21})^{1/2} = 1$; $\lambda_- = \lambda_{11} - \lambda_{22} = 0$; $h = H_{c2}D_1\hbar/2\phi_0 k_B T$; $\eta = D_2/D_1$; $U(x) = \psi(1/2 + x) - \psi(1/2)$, $t = T/T_c$, $\phi_0$ is the magnetic flux quantum, and $D_{1,2}$ are the electronic diffusivities for different FS sheets [40]. We assume that the derivative $dH^c_{c2}(T)/dT = -1.68$ T/K close to $T_c$ is determined by $D_1$ for the band with the highest coupling constant, i.e., $D_1 \gg D_2$ [6], and therefore $D_1$ can be deduced from:

$$D_1 \approx \frac{8\phi_0 k_B}{\pi^2 \hbar |dH_{c2}/dT|} = 1.22 cm^2/\sec \quad (5)$$

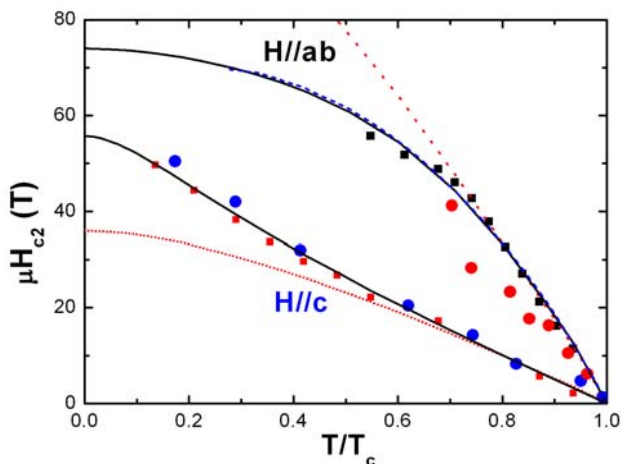

Fig. 5. (Color online) Temperature dependences of the upper critical fields, $H_{c2}(T)$, for KFS1 (circles) and KFS2 (squares) single crystals for $H \parallel c$ (red squares) and $H \parallel ab$ (black squares). The dotted lines indicate the temperature dependences according to the WHH model neglecting Pauli limit. The solid lines show the dependences in a two-band fit by use of Eq. (4) and dashed ones according to Eq.2.

Given this $D_1$ value, the temperature dependence of $H_{c2}(T)$ is accounted for by Eq. (4). The solid line in Fig. 5 for $H // c$ is the best fit of Eq. (4), obtained with $\eta = 0.1$. Therefore, the deduced limiting value of $H_{c2}(0) = 56$ T is likely dominated by a band with low diffusivity $D_2 = 0.12$ $cm^2/sec$, while the slope $dH_{c2}/dT$ close to $T_c$ is due to a band with larger diffusivity, $D_1 = 1.22$ $cm^2$/sec. This two-gap model quantitatively reproduces the unconventional non-WHH temperature dependence of $H_{c2}(T)$ for $H // c$, while the Pauli model works nicely for $H //ab$. Nevertheless, it is not clear why two gap model does not work for $H // ab$ as well?

The overall $H_{c2}(T)$ dependence is in qualitative agreement with earlier data for $K_{0.8}Fe_{1.76}Se_2$ for both field directions [8], even though the reported superconducting transitions for H//ab are very broadened. Unfortunately no Pauli scenario, nor two gap model was treated in this paper for the interpretation of the $H_{c2}(T)$ dependence, which made quantitative comparison difficult. These two scenarios were considered recently for electron doped $Sr_{1-x}Eu_x(Fe_{0.89}Co_{0.11})_2As_2$ [9], with definitely different FS including electron and hole sheets.

Additionally, we did not observed any feature of surface superconductivity for KFS2 samples in contrast to the statement of Ref. [29]. Observation of $H_{c3}(T)$ dependence in Ref. [29] could be due to degradation of the surface layer in air. Indeed, we have observed a drop of $T_c$ from $28$ K to $25$ K, measured by TDO technique after exposing the sample to the air, while larger $T_c$ was restored after subsequent cleavage of the sample. Besides, a much wider transition is observed for KFS1 compared to KFS2 crystal (see data in Fig. 2). This feature is due, not only to a less stoichiometric composition in the former case, hence to a more disordered sample, but also probably to air oxidization of the KFS1 single crystal.

With regards to the anisotropy of the Pauli paramagnetic field, $H_P$ is given by $\mu_B H_p = 1.06 \Delta(0) \eta_{eff}(\lambda)$ [36], where $\eta_{eff}(\lambda)$ describes the effect of gap anisotropy, multi-band character, energy dependence of states etc., and λ is the anisotropic electron-phonon coupling renormalization factor. Thus some anisotropy of $H_p(0)$ we observed in such anisotropic spectra like almost 2D EKFA is not surprising. More surprising is the absence of Pauli paramagnetism for KFS for *H//c*, while it does works at *H//ab*.

We have to mention another point regarding the persistence of magnetic ordering within the superconducting state [24, 25, 26]. Actually, no indication of AFM state suppression can be inferred from our data. Similar conclusion can be derived from the data relevant to electron - doped $Ba(Fe_{0.93}Co_{0.07})_2As_2$ [6] and $Sr_{1-x}Eu_x(Fe_{0.89}Co_{0.11})_2As_2$ single crystals [9], for which the temperature dependence of $H_{c2}$ support a Pauli scenario for *H //ab* and a two gap behavior for *H // c*, as it is the case for KFS. These results raise many questions: (i) To what extent is the AFM in KFS interacting with lower-temperature superconductivity [9]? It has been shown recently [41] that the coexistence of AFM and superconductivity is due to nano scale phase separation between superconducting and AFM grains. Observation of structural lamellae with the Fe-vacancy order and disorder states along the c-axis direction in $K_{0.8}Fe_xSe_2$ single crystals from transmission electron microscopy, is in line with this model [26]. (ii) Why does two-gap fit work for *H // c*, but not perpendicular to this direction? These problems need to be addressed by future careful studies.

## IV. Conclusions

In conclusion, the measurements of $H_{c2}(T)$ for an electron - doped 122 iron-chalcogenide and a hole-doped 122 iron-pnictide superconductors allow to conclude that: (i) for hole-doped EKFA, the temperature dependence of $H_{c2}(T)$ for *H // c* and *H // ab*, is accounted for by the Pauli model, including a slightly anisotropic Pauli-limiting field over the whole temperature region; (ii) for electron-doped KFS, the data support a Pauli scenario for *H//ab* too, while a two-gap behavior is observed for *H//c*; (iii) data are very sensitive to the sample preparation and, likely, to disordering. Air oxidization leads to a rapid degradation of the superconducting properties, namely, a significant decrease of $T_c$ and large broadening of the superconducting transition in magnetic field are observed; (iv) The ratio of the diffusivities for the two-band model in KFS is rather large, $D_1/D_2 = 10$, indicating that the scattering rates of each these bands differ by one order of magnitude, (iv) the coherence length is anisotropic in both compounds but is larger than the thickness of the conducting sheets indicating 3D superconductivity. Despite the coexistence of lattice AFM and superconductivity in both compounds, no influence of magnetic ordering on $H_{c2}(T)$ was observed up to *57 T*.

## Acknowledgments

We would like to thank V.F. Gantmakher and R. Huguenin for helpful discussions. Comments of V.N. Zverev are appreciated too. Part of this work has been supported by EuroMagNET II under the EU contract number 228043, and the German Science Foundation within SPP 1458. This work was supported in part by the Russian Academy of Science Program "Quantum mesoscopic and no homogeneous systems".